\documentclass{article}

\usepackage{graphicx}% Include figure files
\usepackage{dcolumn}% Align table columns on decimal point
\usepackage{bm}% bold math
\usepackage{hyperref}% add hypertext capabilities
\usepackage[mathlines]{lineno}% Enable numbering of text and display math

\usepackage{authblk}

\usepackage{geometry}
\usepackage[minnames=1,maxnames=3,style=numeric,backend=bibtex]{biblatex}
\usepackage{xspace}
\newcommand{\larcv}{\texttt{larcv}\xspace}

\begin{document}

%\preprint{APS/123-QED}

\title{An Efficient, Scalable IO Framework for Sparse Data: larcv3}% Force line breaks with \\

\author[1,*]{Corey Adams}
\author[2]{Kazuhiro Terao}
\author[3]{Marco Del Tutto}
\author[4]{Taritree Wongjirad}
\affil[1]{Argonne National Laboratory}
\affil[2]{SLAC National Laboratory}
\affil[3]{Fermi National Accelerator Laboratory}
\affil[4]{Tufts University}
\affil[*]{Corresponding author: corey.adams@anl.gov}

\date{\today}% It is always \today, today,
\maketitle

\begin{abstract}
Neutrino physics is one of the fundamental areas of research into the origins and properties of the Universe.  Many experimental neutrino projects use sophisticated detectors to observe properties of these particles, and have turned to deep learning and artificial intelligence techniques to analyze their data.  From this, we have developed \larcv, a \texttt{C++} and \texttt{Python} based framework for efficient IO of sparse data with particle physics applications in mind.  We describe in this paper the \larcv framework and some benchmark IO performance tests.  \larcv is designed to enable fast and efficient IO of ragged and irregular data, at scale on modern HPC systems, and is compatible with the most popular open source data analysis tools in the Python ecosystem.

\end{abstract}

%\tableofcontents

\section{\label{sec:intro}Introduction}

In neutrino physics, high-resolution imaging detectors are driving the scientific advances of experimental programs.  Reconstruction of this data is challenging, and Artificial Intelligence and Machine Learning are some of the leading techniques for reconstruction and analysis of neutrino physics detectors \cite{uboone,next,sbnd,sparse-data}.

Neutrino data, unlike traditional image data, is typically sparse and irregular. The native resolution of the detector data, disregarding sparsity, is usually significantly higher than industry-standard examples such as ImageNet \cite{imagenet}. Additionally, labels for datasets require flexibility to store a variety of information: whole-event labels, segmentation masks, instance masks, as well as physical properties like particle energy and locations.  The size of detector images, typically O(1000) pixels on an axis and with 2 or 3 dimensions, necessitates either coercing traditional image storage techniques to accommodate large images with flexible labeling schemes, or developing a novel storage solution that directly leverages the inherent sparsity of this data.  In this paper we describe a storage, retrieval, and processing system, built on \texttt{HDF5} \cite{hdf5} in \texttt{C++}, that is capable of high-performance IO for neutrino physics datasets for machine learning training and inference.  This framework is designed to work natively at all scales from personal laptops to small workstations to the largest HPC systems available.  Due to its origin from the community of ``\textbf{L}iquid \textbf{Ar}gon'' experiments, and it's focus on \textbf{C}omputer \textbf{V}ision applications, we call this framework ``\larcv''.

In this paper, we present first the landscape in which \larcv was created as well as the niche it is expected to fill, working as a tool in collaboration with other frameworks rather than reinventing any already-successful software.  In Section \ref{sec:datamodel}, we provide an overview of the data schema and conceptual model of data products that \larcv works with, including some details of the serialization model as relevant.  Finally, Section \ref{sec:performance} presents performance results of this framework from real use cases.

\section{\label{sec:motivation} Use Cases, Motivation, and Related Work}

\larcv is focused on AI and data analysis use cases for particle physics and ``big data'' fundamental physics experiments.  Example use cases include loading large batches of example training data in an AI training application, potentially across multiple nodes; or, applying on-the-fly augmentation and pre-processing of data, in sparse format, to make the most of a data set without needing to convert to dense formats.  \larcv is also deliberately designed to be ``experiment-agnostic'' in that there is no reference to any particular experimental apparatus, or its datatype, in the framework.  Instead, information necessary to interpret the dataset(s) is embedded directly into the files as flexible and descriptive metadata.

The implementation of data models for particle physics data is not new. On the contrary, it is a common step for nearly all particle physics experiments that aim to serialize and then reload data for offline analysis.  Each data model tends to be unique to an experiment, as well as deeply tied to a computing framework such as \texttt{art} \cite{art}.  Because of this, the development of AI methods and tools is largely compartmentalized into experimental efforts that share, at best, high-level concepts and results.

Developed by the particle physics community, the \texttt{larsoft} framework \cite{larsoft} is a dedicated simulation, analysis, and data processing framework for high energy neutrino physics experiments such as DUNE \cite{dune} and the Short Baseline Neutrino program \cite{sbn}.  Though the flexibility offered by \texttt{larsoft} is sufficient for the storage of datatypes of a variety of experiments, it suffers from a number of deficits that prevent its direct use in modern data science and artificial intelligence frameworks.  Firstly, \texttt{larsoft} is pure \texttt{C++}, and attempts to expose it to \texttt{Python} never became widely adopted.  Secondly, it is built upon the \texttt{ROOT} framework \cite{root} which - while advantageous for many particle physics applications - is not well suited to parallel IO and HPC environments.  Thirdly, the installation and platform support for \texttt{larsoft} is focused almost entirely on Scientific Linux for data processing support of neutrino experiments.  It is poorly suited to AI applications and HPC environments, though notably, some frameworks such as \texttt{tensorflow} and \texttt{pytorch} could integrate with \texttt{larsoft} via their \texttt{C++} APIs, enabling model inference with only moderate hardship.

On the other hand, efficient and well-support data loading models for AI applications exist, such as \texttt{tensorflow}'s \texttt{TF.Dataset} and \texttt{pytorch}'s \texttt{DataLoader}.  Further, extensions such as \texttt{pytorch\_geometric} implement data models designed to handle irregular data similar to those found in particle physics experiments.  Despite appearances, none of these are well suited to the sparse data in particle- and neutrino-physics experiments.  The \texttt{pytorch} and \texttt{tensorflow} data interfaces are highly focused on dense image or time-series data, as driven by industrial applications.  \texttt{pytorch\_geometric} is focused on dynamic graph applications, while particle physics datasets are typically highly sparse but embedded in a voxelized, regular space.

With all of this in mind, the main design goals of \larcv can be summarized as:
\begin{itemize}
    \item \larcv must provide a way to retrieve data from particle physics frameworks, store it to disk, and enable fast retrieval (potentially at scale) for model training and development.
    \item The implementation must be agnostic to particular experimental details but sufficiently descriptive and flexible to be broadly applicable.
    \item The interface to the data model must be fully available both in \texttt{C++} and \texttt{Python} for full support of conversion of data from bespoke, experimental formats (i.e., \texttt{larsoft}) to the more generic \larcv format.
    \item The  \larcv framework must be built upon common, open source tools that are easily installed or build.  In this case, we use  \texttt{HDF5} \cite{hdf5} for IO,  \texttt{json} \cite{json} for configuration, and \texttt{pybind11} \cite{pybind11} plus \texttt{NumPy} \cite{numpy} for exposing data directly to \texttt{Python} and thus the Python ecosystem.
    \item The \larcv framework must be intrinsically scalable and performant in particular for reading of data during an analysis task, either on a single system or on a large distributed system.
\end{itemize}

\section{ \label{sec:datamodel} Data Model}

To enable efficient IO within \larcv, we have designed a data model where users will format their data to fit into several abstract, optimized data products which can then be efficiently and automatically serialized or de-serialized.  With \texttt{HDF5} \cite{hdf5} as a backend for IO, we take advantage of the high performance, open-source and cross-platform file formats.  We build several data types on top of this to accommodate the unique needs of sparse, high energy physics data.  Additionally, \texttt{HDF5} enables \larcv to handle sparse IO at scale with MPI-IO as needed, and \larcv is well prepared to take advantage of data-store systems as they come online in future HPC systems.

Particle physics data is typically multi-channeled with several sources of data per single observed (or simulated) interaction.  The atomic unit of a particle physics data-set is often called an ``event'' and represents one self-contained observation (or simulation) of a detector.  An event can contain images from an imaging detector, segmentation labels from a simulation, and multiple projections of a single detector's readout (as in a LArTPC \cite{uboone, sbnd}).  Additionally, simulated events typically contain a list of ``particle'' objects representing the trajectories of fundamental particles, their energies, and their hierarchy of which particles created which other particles.  It is a fundamental design principle of \larcv to enable efficient storage and use of these pieces.

\begin{figure}[ht!]
    \centering
    \includegraphics[width=0.45\columnwidth]{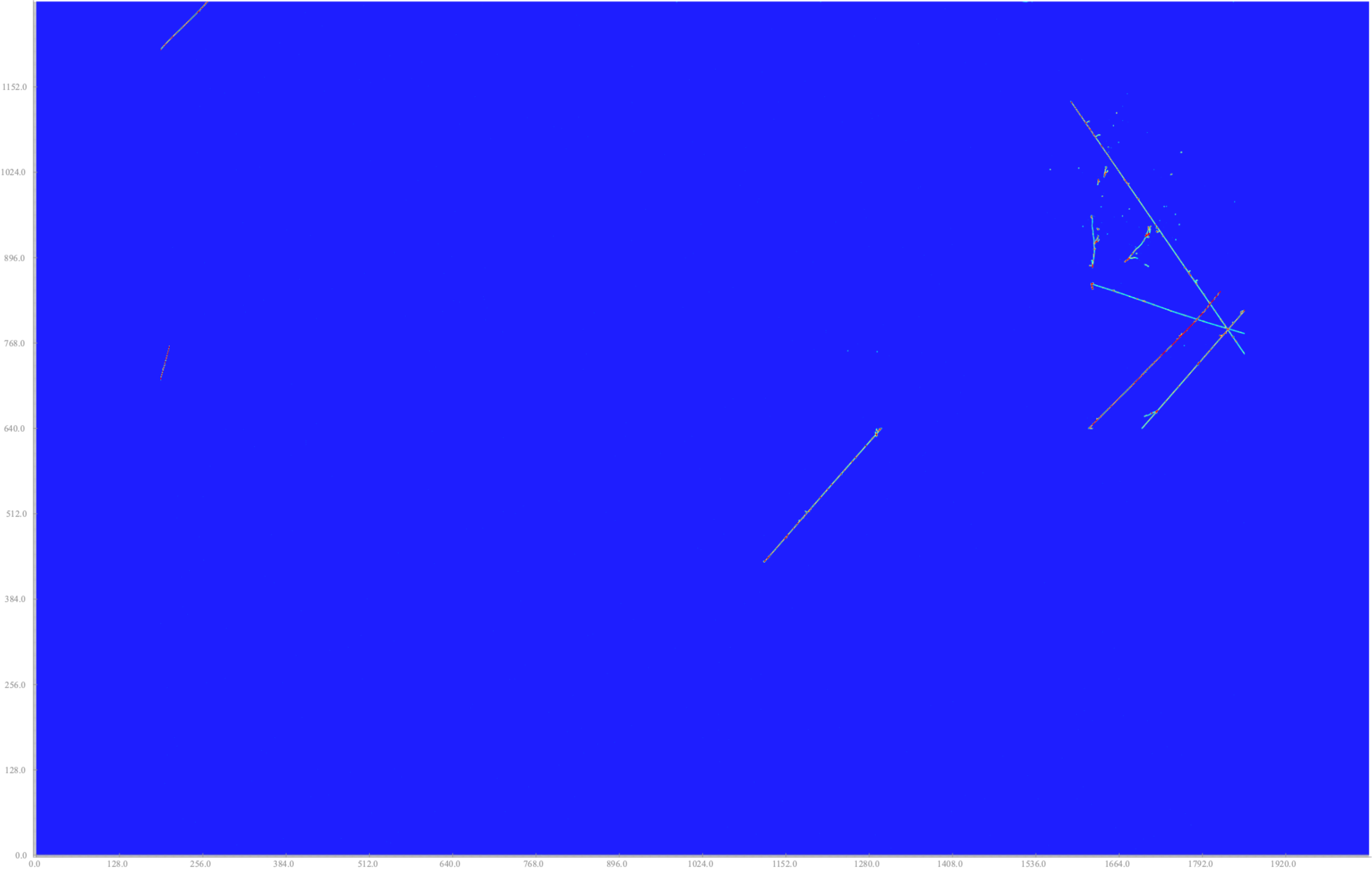}
    \includegraphics[width=0.45\columnwidth]{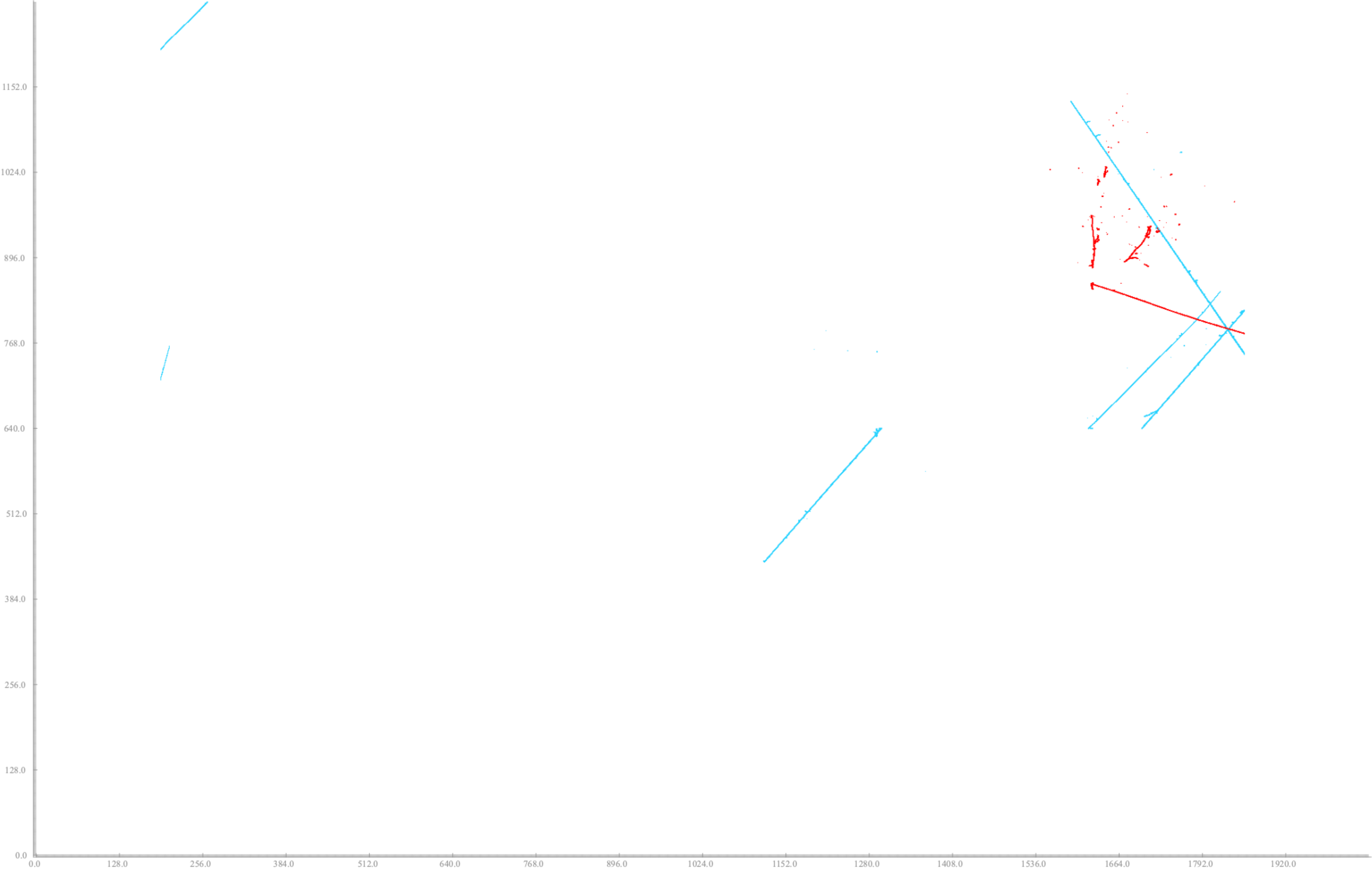}
    \caption{Examples of 2D data, from the CosmicTagger dataset (described below).  Left is a dense representation of simulated input data, with an artificial blue background.  Right is the sparse representation of the corresponding segmentation labels.}
    \label{fig:CT_wire_segmentation}
\end{figure}

\begin{figure}[ht!]
    \centering
    \includegraphics[width=0.45\columnwidth]{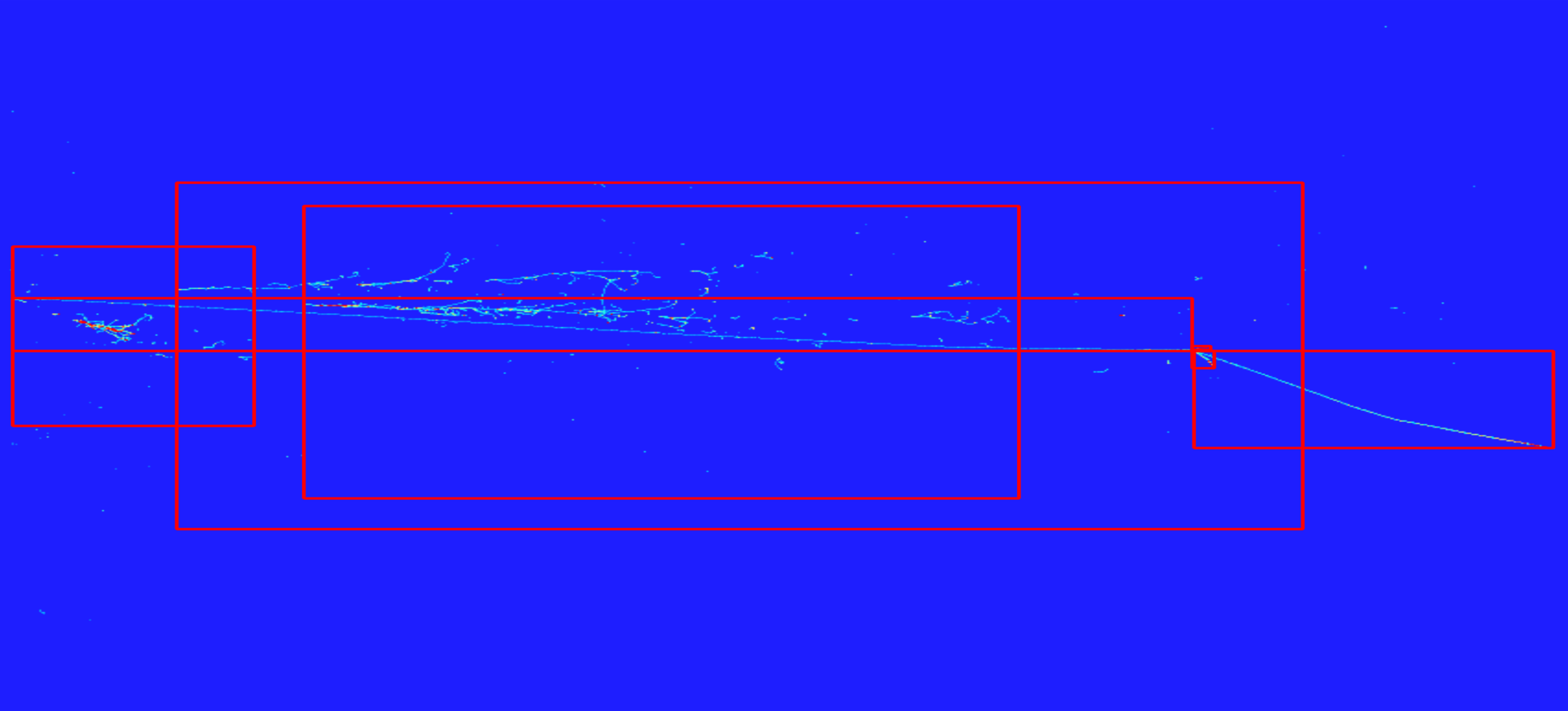}
    \includegraphics[width=0.45\columnwidth]{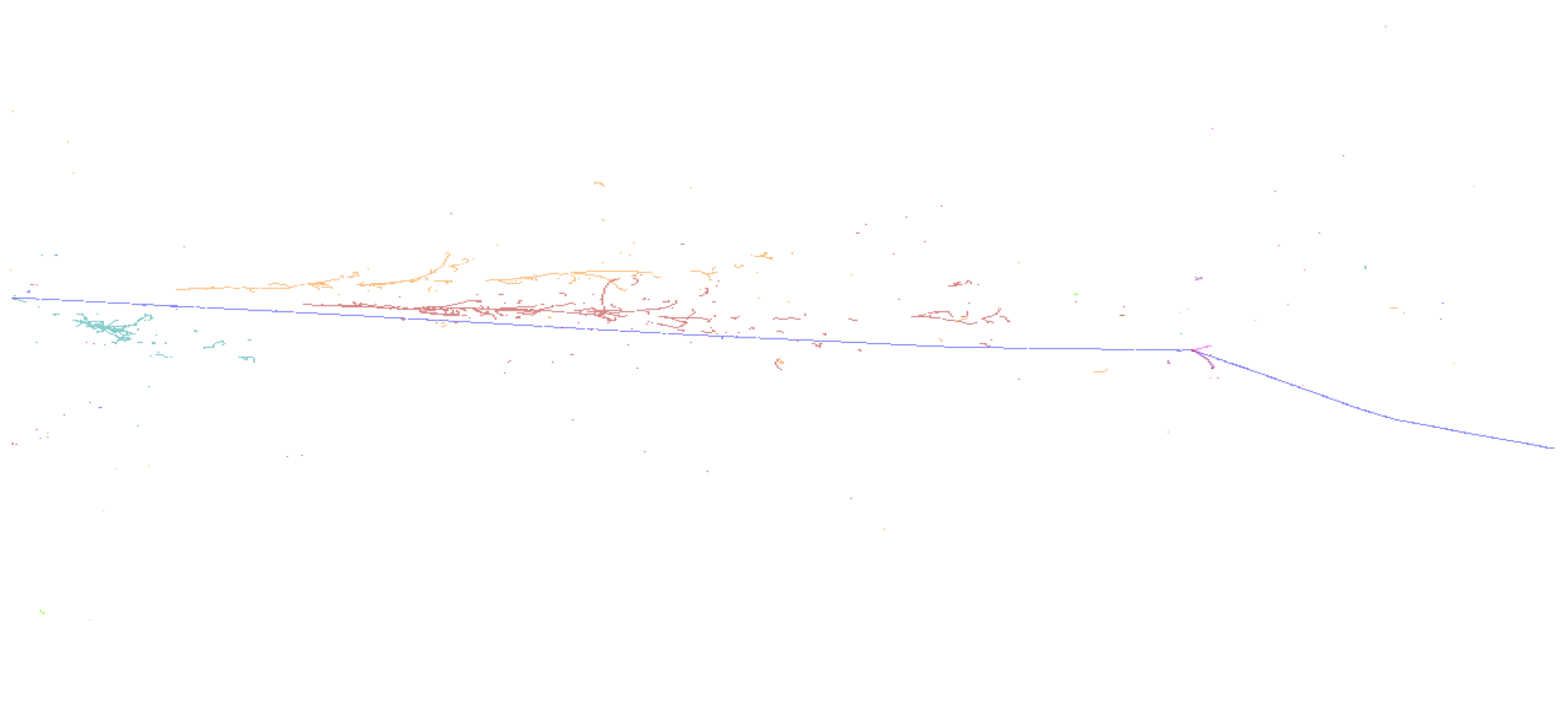}
    \caption{Examples of 2D data, from the DUNE2D dataset (described below).  Left is a dense representation of simulated data, with an artificial blue background, overlaid with instance bounding boxes.  Right is the SparseCluster representation of the corresponding instance segmentation labels.}
    \label{fig:dune2d_data}
\end{figure}

\begin{figure}[ht!]
    \centering
    \includegraphics[angle=-90,origin=c,width=0.45\columnwidth]{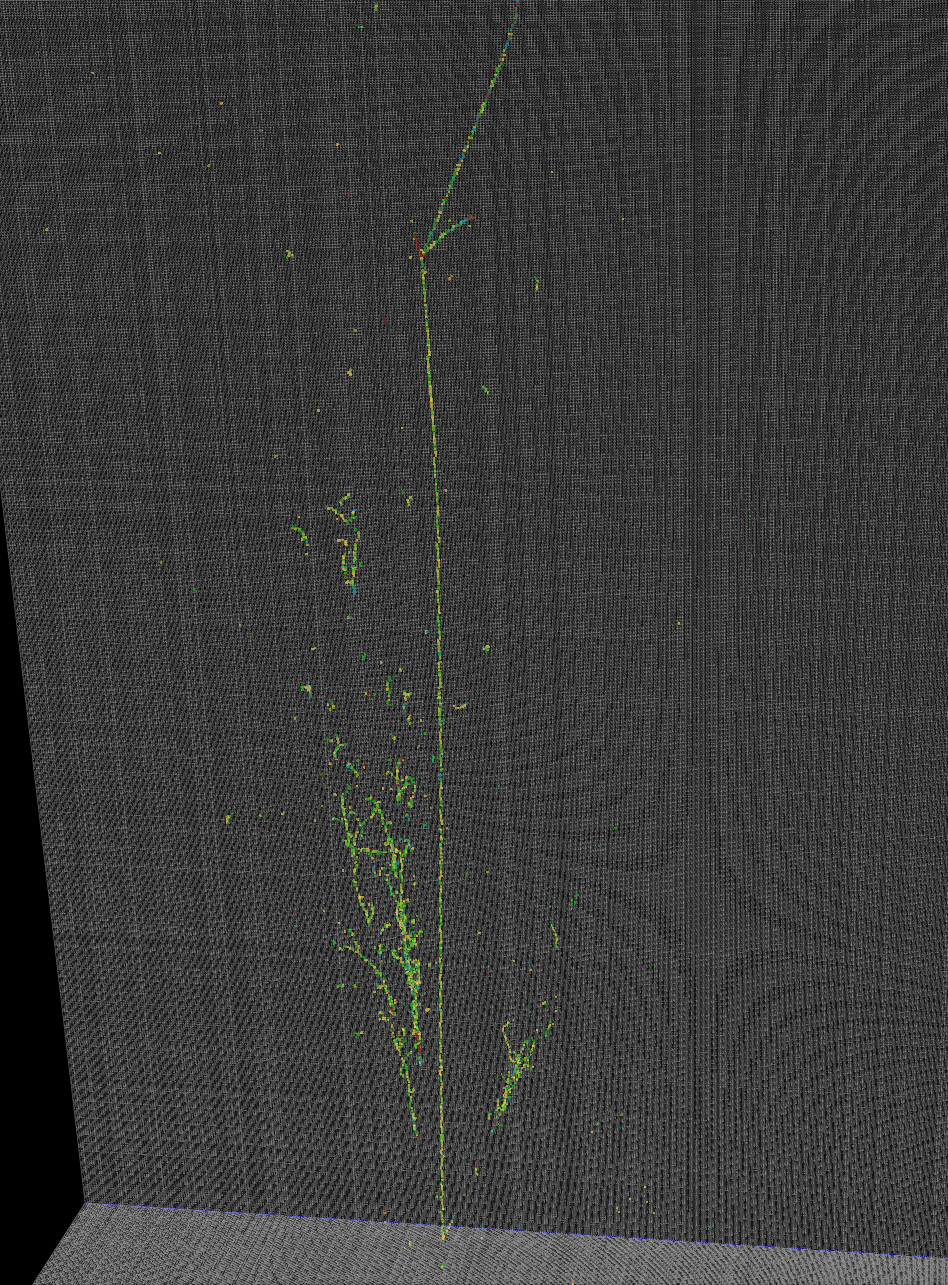}
    \includegraphics[angle=90,origin=c,width=0.45\columnwidth]{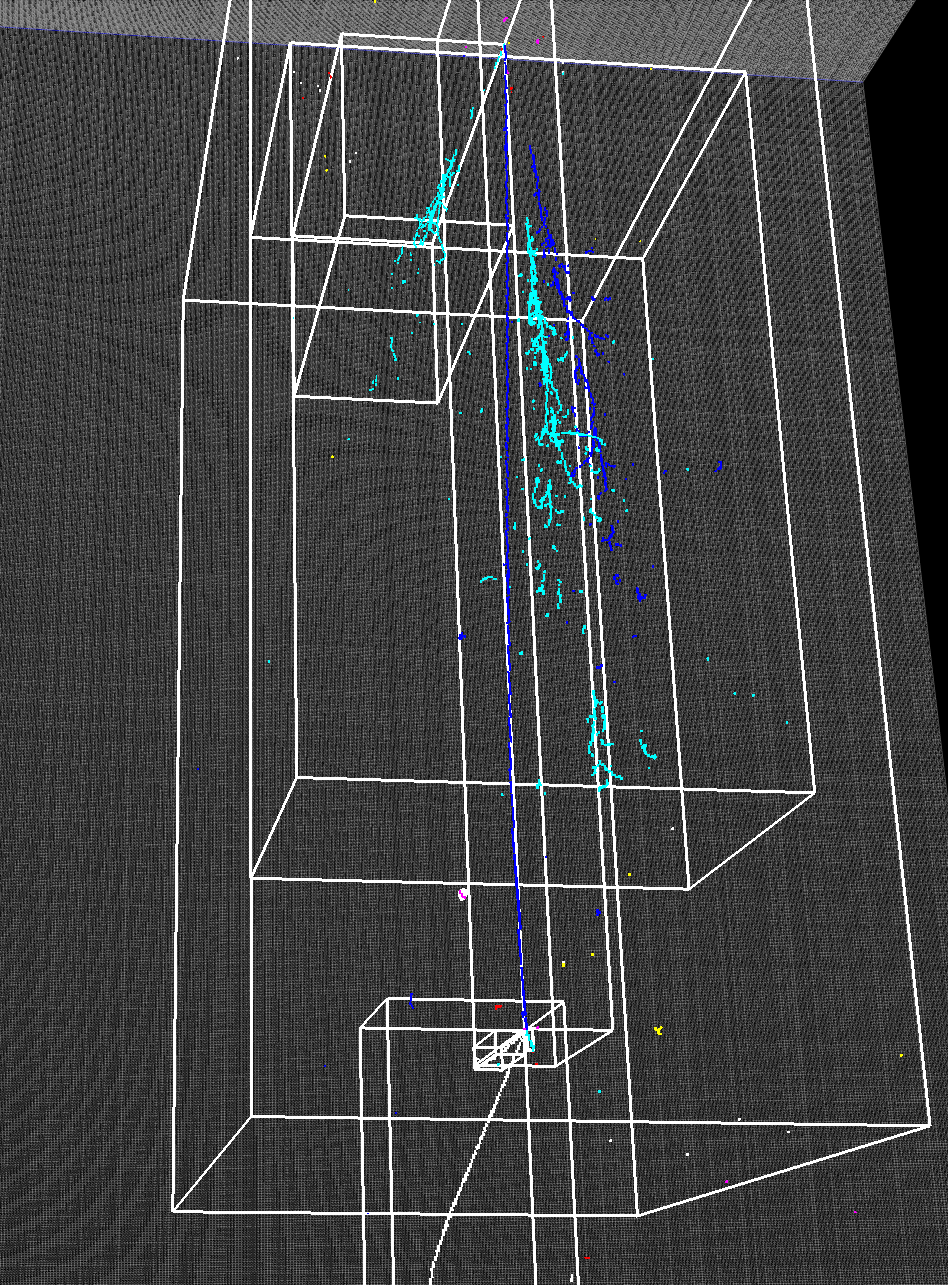}
    \caption{Examples of 3D data, from the DUNE3D dataset (described below).  Left is a dense representation of simulated data in 3D.  Right is the SparseCluster representation of the corresponding instance segmentation labels, shown with instance bounding boxes overlaid.}
    \label{fig:dune3d_data}
\end{figure}

The fundamental data objects of \larcv are as follows
\begin{enumerate}
    \item \textbf{Tensors (1D, 2D, 3D or 4D)} - A tensor is a dense, memory-contiguous object that can be interpreted as a multi-dimensional array.  For better integration into the Python ecosystem, the indexing scheme adopted by \larcv to map from contiguous memory location to N-dimensional index is deliberately identical to the scheme used by \texttt{NumPy} \cite{numpy}.  Once deserialized from disk, a Tensor in \larcv can be yielded to \texttt{NumPy} with zero data copy.  Examples of Tensor2D data are shown in Figures \ref{fig:CT_wire_segmentation} and \ref{fig:dune2d_data}.
    \item \textbf{SparseTensor (2D or 3D)} - A sparse tensor is a zero-suppressed implementation of a Tensor object, where the key data encoded is a ``Voxel'' object consisting of a (value, index) pair.  The index represents the index of this voxel if it were stored in a contiguous-memory tensor, while the value is self-explanatory. Sparse tensors are zero-suppressed in concept, but it is permissible to deliberately store a 0.0 value if that is needed by an application.  SparseTensors can represent the intrinsic data of an experiment, or derived objects such as sparse segmentation masks.  SparseTensor data is shown in Figure~\ref{fig:CT_wire_segmentation}.
    \item \textbf{SparseCluster (2D or 3D)}- A sparse cluster is an extension of a sparse tensor where each voxel is effectively given a third property, an index of a ``cluster'' to which it belongs.  In particular, this is targeting instance segmentation tasks where multiple masks are part of the labeled data set for any given image.  SparseCluster data is shown in Figures~\ref{fig:dune2d_data} and \ref{fig:dune3d_data}.
    \item \textbf{Particle} - A particle represents high-level simulated or reconstructed information about an Event as a whole, or a list of particles can represent the detailed information about every particle tracked in an event.  Particle objects contain, for example, energy, vertex, and momentum information.  They also may be used for entire-event classification storage.  Particles can store hierarchy information, and can also be used to create hierarchies of associated objects (SparseClusters, Bounding Boxes).
    \item \textbf{Bounding Box  (2D or 3D)} - A rectangular object that dictates the extent of an object of interest in a detector.  Similar to the bounding boxes of object detection work in computer vision \cite{faster-rcnn}, though extended with rotation matrices to enable a broad suite of use cases (lines, rectangles, at angles, etc.).  Bounding box example data is show in Figures~\ref{fig:dune2d_data} and \ref{fig:dune3d_data}.
    \item \textbf{ImageMeta (1D, 2D, 3D or 4D)} - A meta class describes the voxelization of an N-dimensional space, including the location of the origin, size, and number of voxels per dimension.  While not a dataproduct in and of itself, the ImageMeta class provides the context for the spatial dataproducts and therefore also defines how they interact.
\end{enumerate}

With the exception of the Particle data product, each of the data products has an associated vector of ``projection IDs'' which can serve as a multipurpose enumeration of detectors: it can be the multiple wire-planes of a LArTPC, or several readout regions in a segmented detector.  It is up the creator of the data to decide.  To describe the properties of the volume indicated by the projection ID, an ImageMeta object is stored with each projection that describes the physical volume in terms of number of voxels, voxel size, and origin across all dimensions.  The ImageMeta objects also provide a convenient interface for raveling/unraveling a multi-dimension index, identical to the \texttt{NumPy} scheme of mapping a multi-index to a single flattened array index.  The ImageMeta interface is shared across all data products that have a spatial relation, ensuring consistent development patterns.  The ImageMeta objects also provide the absolute positioning of all objects, enabling full images to be built from objects that represent only a subset of a detector, potentially.

\subsection{Serialization Methods}

Each larcv file is divided into two H5Groups: {\bf Events/} and {\bf Data/}.  Within {\bf Events/} there is only one table, {\bf event\_id}, which enumerates the global bookkeeping properties of each event in the file.  Events may be stored in any order, and are tagged through a run/subrun/event indexing scheme.  A run is usually an index that is constant through one data-taking or simulation-generating unit of a dataset, where detector conditions are constant, as dictated by a particular experiment.  A subrun frequently indicates a sub-file within a run, and events indicate an entry index within a file.  Though \larcv is based on real-world situations with run/subrun/event indexing, these are purely informational objects in \larcv.

All data products are stored in a product/producer pair: there is no particular limit on the number of a particular data product stored, but it must have a unique producer name.  Further, the data product for each producer is stored for every event.  To enable a complete implementation of sparse and dense tensors, as well as sparse cluster sets and particle arrays, \larcv uses a mix of inheritance and dimensional templating to have a minimal code base for both serialization/deserialization and python bindings.  Python binding implementation is done via pybind11 \cite{pybind11}, while configuration is done via the \texttt{nlomann\_json} package \cite{nlohmann, json}.

% Each dataproduct in \larcv is initially stored in local memory, with buffers flushed at the transition between event indexes.  In some cases, such as on-the-fly pre-processing, it is advantageous to allow creation of data products in memory without serialization to disk.  \larcv will automatically enable this in, for example, read-only mode.  Alternatively, the user can have full, fine-grained control of which data products are written (and read) from disk through the configuration files.

\subsection{De-serialization, Pre-processing and Data Loading}

\larcv uses an on-demand deserialization procedure, where no data is loaded until it is requested from file.  A dedicated class, \texttt{IOManager}, keeps track of the current Event index as well as what objects are loaded from disk.  When a user requests a data product, with an associated producer label, \texttt{IOManager} will load the entire object for the current Event index from disk and store it in memory.  Subsequent calls return the in-memory objects without disk access until processing of the current Event is finished.

\larcv features a dynamic object storage system where users can create data product objects on-the-fly during runtime, as well as apply augmentation techniques to data products.  This is useful in particular for applying augmentation techniques for training deep neural networks, as the typical transforms (mirroring, rotations, etc) do not have easily accessible sparse algorithms available.

For training a neural network or similar batch-processing based workloads, a special interface to \larcv exists: the \texttt{QueueIO} class. This provides queued, in-the-background, data loading as well as an easy interface to the data products from \texttt{Python}.  In this mode, \larcv will start loading the next data from disk, and apply any desired processing, while the user's application is still processing the currently loaded data.  For GPU-based applications, this provides excellent overlap of CPU-based IO with GPU-based computing, and can reduce IO-related time to negligible levels.  The implementation relies on \texttt{std::future} objects in \texttt{C++}.  Additionally, the queued IO implementation is intrinsically scalable, as shown in Section~\ref{sec:performance}.  Users may dynamically configure the indexes of the next batch, or \larcv can randomize the dataset on the fly.  Sequential access is available as well, well suited for inference modes, and a hybrid ``random-block'' interface draws locally sequential, globally random sequences of events to improve disk access patterns without fully sacrificing dataset randomization.

\subsection{Visualization Tool}

Due to the unique data format and serialization scheme of \larcv dataproducts, we provide a python-based visualization tool, simply called \texttt{larcv-viewer} and available publically on GitHub \cite{larcv-viewer}. The viewer is also based on open-source tools, particularly \texttt{PyQtGraph} \cite{pyqtgraph} and \texttt{PyQt5} \cite{PyQt5}.  All images of data shown in this paper were produced with this visualization tool.

\section{\label{sec:queue_io} Queued IO with MPI}

To demonstrate the performance of \larcv on real applications, we benchmark \texttt{QueuedIO} performance and scalability with MPI on two published datasets.  The first dataset, referred to as ``CosmicTagger'' comes from the Short Baseline Near Detector and is used in a segmentation application, described here \cite{cosmic_tagger}.  The CosmicTagger dataset is focused on 2D image analysis, featuring three images per event of resolution $1280 \times 2048$.  The CosmicTagger dataset also features the data stored in both sparse and dense format, on disk, to enable a comparison.

The second dataset is referred to as ``DUNE'' and is described in \cite{pixwire}.  This dataset contains both 2D and 3D images, where the 2D images are projections of the 3D data onto three different ``views'' (similar to the three images in the CosmicTagger dataset).  The details of the datasets are described in Table~\ref{tab:datasets}.

\begin{table}[ht!]
\centering
\begin{tabular}{r|ccccc}
Dataset & Dim. & Image Size & Events& $\bar{N}_{non-zero}$ & Occupancy \\
\hline \hline
Dune  & 2D & (1536,1024,3)  & 5854981  & 4175 & 8.85E-04  \\
Dune & 3D & (900,500,1250)  & 5854981  & 1921 & 3.41E-06 \\
CosmicTagger & 2D & (1280,2048,3) & 43075 & 57138 & 7.27E-03  \\
\end{tabular}
\caption{Meta data of datasets used in this benchmarking paper.}
\label{tab:datasets}
\end{table}

Both datasets feature sparsity to a large degree, more so than typical dense computer vision datasets.  A notable difference is that the CosmicTagger dataset features a noise model as well as neutrino simulation effects, with more pixels non-zero than either version of the DUNE dataset.  To showcase the relative performance of sparse vs. dense IO, the images in the cosmic tagger dataset are available in both \texttt{Tensor2D} and \texttt{SparseTensor2D} format.  \larcv can read dense data to a dense output \texttt{NumPy} array, while sparse data may be read to either dense or sparse output arrays.  The ``sparse'' \texttt{NumPy} array is similar to the format suitable for a Submanifold Sparse Convolutional Network \cite{scn}, popular in AI applications of neutrino physics.

\begin{figure}
    \centering
    \includegraphics[width=\textwidth]{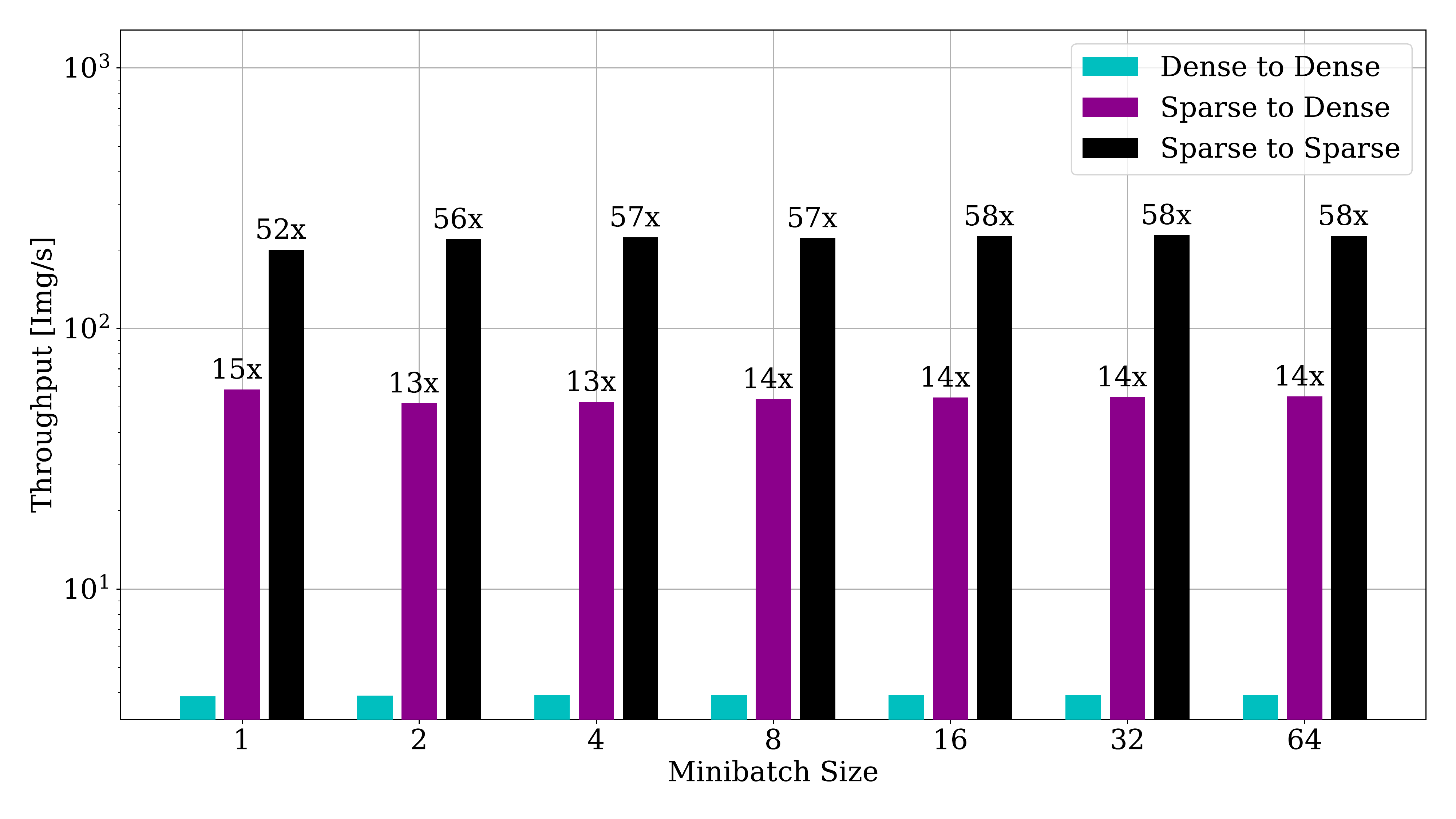}
    \caption{Comparison of IO throughput for the cosmic tagger dataset, as a function of minibatch size, for the possible input and output data shapes. The numbers above the bars are with respect to the dense-to-dense configuration.}
    \label{fig:ct_dataformat}
\end{figure}

Figure~\ref{fig:ct_dataformat} shows the comparative performance, on a single process, of \larcv reading the various possible configurations of CosmicTagger's data.  The dense-to-dense dataset represents reading dense, though compressed, tensor data from disk and yielding dense \texttt{NumPy} arrays to python.  Using sparse data on disk, instead of dense data, produces a 13x to 15x image throughput for the same problem size, while using sparse output instead of dense output shows a $>$50x performance increase from the original configuration.  From this, particularly for GPU-based applications where data must be fed from CPU to GPU, we recommend reading data to sparse format and expanding to dense format - if needed - directly in GPU memory to reduce data movement latency.  In Section~\ref{sec:performance}, we use sparse input and output for all scaling benchmarks but note that scaling performance is approximately equal in all data formats.
\section{\label{sec:performance} Performance Benchmarks }

The main expected use case of \larcv is fast, parallel, and scalable IO for Deep Learning applications, including training and inference.  We have several benchmark comparisons presented here, representing the most common and important use cases of \larcv's sparse IO.  This includes ability to load large minibatches of data in a single process, as well as distribute that work within a node or out onto a large-scale cluster.  All tests were performed on the Polaris \cite{polaris} system, consisting of 512+ nodes of AMD Milan CPUs + 4 Nvidia A100 GPUs.  The GPUs were not used in these tests - we reserve the GPUs for compute-intensive applications and expect IO to happen in parallel on the CPU.

\subsection{\label{sec:single_perf} Single-Node Performance }

\begin{figure}
    \centering
    \includegraphics[width=\columnwidth]{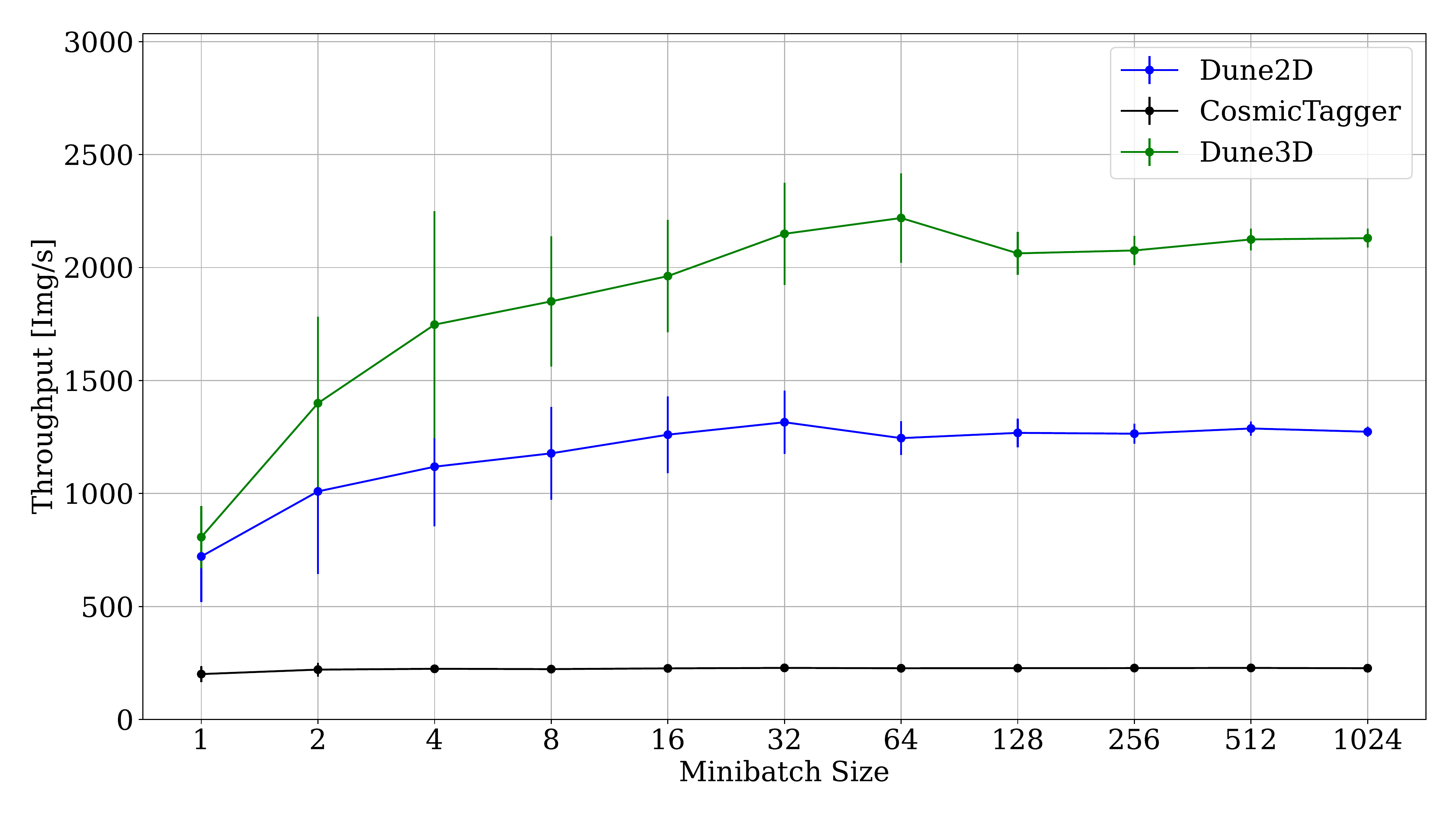}
    \caption{Single process performance of \larcv as a function of minibatch size for all three datasets, using sparse input and sparse output data.}
    \label{fig:single_process}
\end{figure}

A user running a Deep Learning training application will frequently want to saturate GPU memory with each minibatch.  In this use case, we compare the saturation curve of \larcv on a single node, without MPI-parallelization.  In Figure~\ref{fig:single_process}, we see the batch-size scale-up performance of \texttt{larcv} on the benchmark datasets.  In all cases, the throughput increases approximately linearly with batch-size up to a point, and then throughput is approximately flat regardless of batch-size.  Each measurement is the average value from a set of 50 runs.  Because the level of sparsity is variable on an event-by-event basis, the standard deviation of throughput appears particularly high in small-batch instances for the Dune datasets.  We note that in this case, the latency per batch is measured in milliseconds, and the variations are a product of the dataset itself.  Larger batch sizes require total disk usage that effectively averages the dataset variations.

An additional benchmark that is critical for measuring single-node performance is the scale-up with respect to the number of ranks per node.  Most high performance computing nodes are heterogeneous compute nodes, with one or two CPUs in conjunction with multiple GPUs.  Practical applications will run one rank per GPU, however, with some smaller applications and modern, large-memory GPUs, more ranks per GPU can be advantageous.  This is true in particular for low-latency inference using, for example, MPS- or MIG-mode on Nvidia GPUs \cite{mps} where multiple ranks per GPU can bring increases in throughput.  An additional use case is training a Sparse Convolutional Network \cite{scn}, which is a CPU-GPU hybrid framework that will saturate CPU usage at relatively low batch-size.  In this case, multiple small-batch ranks will outperform a single large-batch process by increasing parallelization on the CPU, rather than the GPU.

\begin{figure}
    \centering
    \includegraphics[width=\columnwidth]{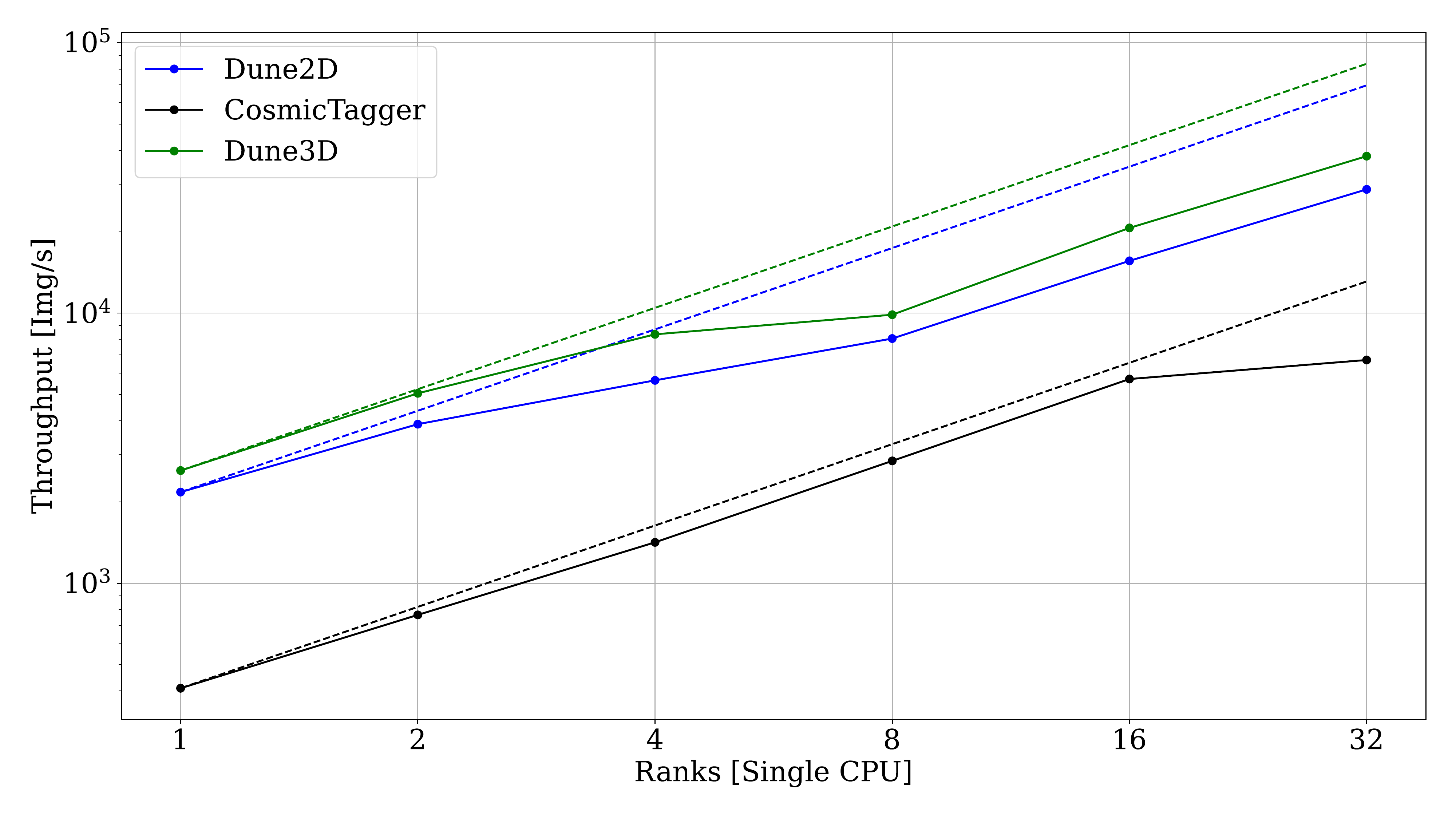}
    \includegraphics[width=\columnwidth]{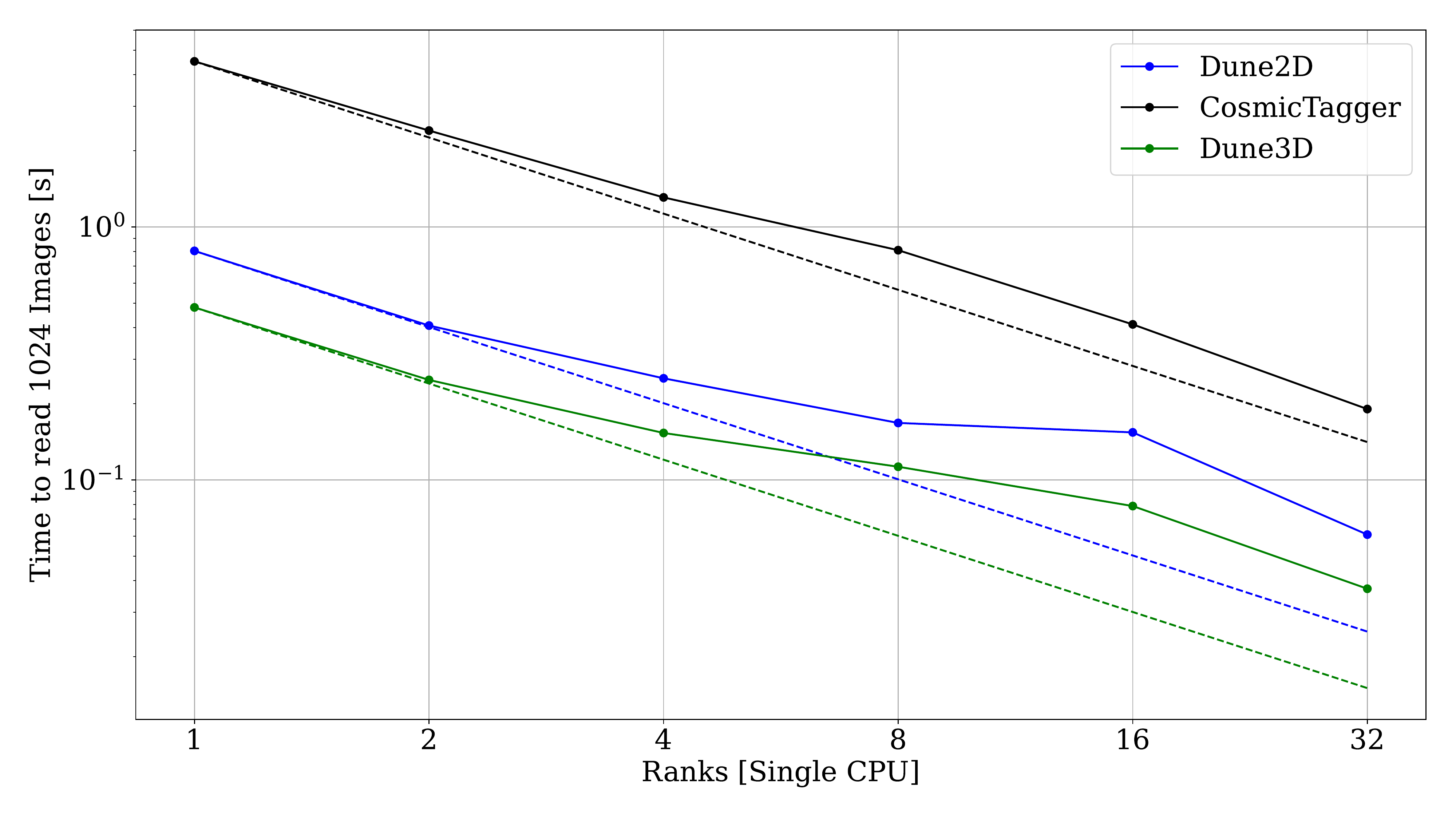}
    \caption{Weak (top) and Strong (bottom) scaling of \larcv benchmarks on a single CPU.}
    \label{fig:single_node}
\end{figure}

Figure~\ref{fig:single_node} shows the single-node scale out of \larcv using MPI for both weak and strong scaling.  The problem sizes are fixed at minibatch sizes of 1024 for all datasets at the largest problem size.  For weak scaling, each rank reads 32 images per rank, up to 32 ranks.  In the strong scaling case, the total images read is fixed at 1024 images, with the work shared by 1 to 32 ranks.  Some degradation is seen as the number of ranks increases, though notably the IO latency remains small compared to the GPU compute time of the applications using these datasets, and overlapping CPU/GPU processing yields negligible IO latency with multiple ranks per node.

\subsection{\label{sec:multi_perf} Multi-Node Performance }

\begin{figure}
    \centering
    \includegraphics[width=\columnwidth]{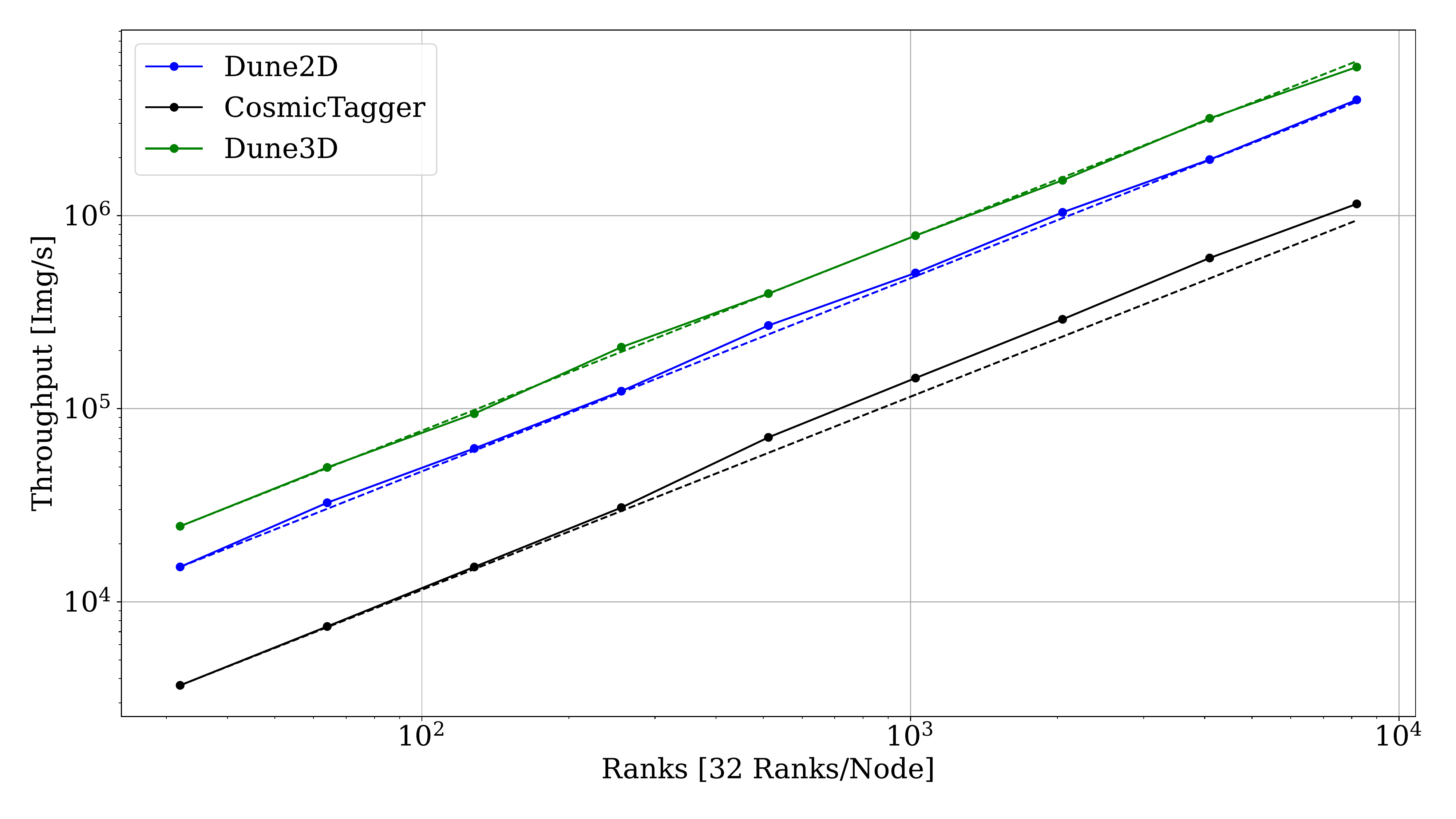}
    \caption{Multi-node scaling of selected datasets with \larcv.  All datasets are reading from sparse format on disk to sparse format in local memory.  Nearly ideal weak scaling is shown here up to 512 nodes, 32 ranks per node, on the Polaris system at ALCF.}
    \label{fig:multinode}
\end{figure}

Finally, the weak scaling efficiency of \larcv is shown in Figure~\ref{fig:multinode} for all three datasets. In this case, the work per CPU is fixed (with 32 ranks per node) and the total batch size is increased.  Notably, above 16 nodes (512 ranks), the total amount of data in CosmicTagger exceeds the total dataset size.  We hypothesize that optimizations in data access patterns of the storage system yield to faster access to data that has already been read, leading to better-than-ideal scaling in the second epoch through the dataset.  Otherwise, we find linear scaling up to the entire system of 512 nodes (16,384 ranks) though note that due to the sparsity, total bandwidth is still below the limits of the lustre parallel file system.

\section{\label{sec:conclusion} Conclusions}

We have presented \larcv, a scalable and efficient framework targeting sparse IO of irregular datasets, particularly from high energy physics.  The software is available open source, used in a variety of applications, cross platform, and accessible in both \texttt{C++} and \texttt{Python}.  Built exclusively on industry standard software such as \texttt{HDF5}, \larcv is compatible with many applications.  We foresee future development in several areas, particularly pre-loading data to accelerators and improving coverage of pre-processing techniques viable for neutrino datasets.  Additionally, direct support of irregular graph data is expected in the future.

% \bibliographystyle{unsrt}
% \bibliography{main.bib}
\printbibliography[title={References}]
\end{document}